\documentclass{pasa}%

\usepackage{graphicx,url,amssymb,amsmath,rotating,color,units,wasysym,epsfig,multirow,epstopdf}
\usepackage[colorlinks,urlcolor=blue,citecolor=blue,linkcolor=blue]{hyperref}

\usepackage{soul,xcolor}
\usepackage{aas_macros}

\jid{PASA}
\doi{10.1017/pas.\the\year.xxx}
\jyear{\the\year}

\newcommand{\NEMO}{NEMO}

\title[Neutron Star Extreme Matter Observatory]{
Neutron Star Extreme Matter Observatory:
A kilohertz-band gravitational-wave detector in the global network\vspace{-1.8cm}}

\author[Ackley, Adya, Agrawal et al.]{K. Ackley$^1$, V. B. Adya$^2$, P. Agrawal$^3$, P. Altin$^2$, G. Ashton$^1$, M. Bailes$^3$, E. Baltinas$^4$, A. Barbuio$^3$, D. Beniwal$^5$, C. Blair$^4$
% ...

\affil{$^1$OzGrav, School of Physics and Astronomy, Monash University, Clayton VIC 3800, Australia}

\affil{$^2$OzGrav, ANU Centre for Gravitational Astrophysics, Research Schools of Physics, and Astronomy and Astrophysics, The Australian National University, Canberra, 2601, Australia}

\affil{$^3$OzGrav, Centre for Astrophysics and Supercomputing, Swinburne University of Technology, Hawthorn VIC 3122, Australia}

\affil{$^4$OzGrav, University of Western Australia, Crawley, Western Australia 6009, Australia}

\affil{$^5$OzGrav, University of Adelaide, Adelaide, South Australia 5005, Australia}

\affil{$^6$The Chinese University of Hong Kong, Shatin, NT, Hong Kong}

\affil{$^7$OzGrav, School of Physics, University of Melbourne, Parkville, Victoria 3010, Australia}

\affil{$^8$Charles Sturt University, Wagga Wagga, New South Wales 2678, Australia}

\affil{$^9$Sydney Institute for Astronomy, School of Physics, University of Sydney, NSW 2006, Australia}

\affil{$^{10}$ATNF, CSIRO Astronomy and Space Science, PO Box 76, Epping, NSW 1710, Australia}

\affil{$^{11}$Macquarie University, Department of Physics and Astronomy, Sydney, Australia}

\affil{$^{12}$Griffith University, Centre for Quantum Dynamics, Brisbane, Australia}

\affil{$^{13}$University of Birmingham, Birmingham B15 2TT, United Kingdom}

\affil{$^{14}$Niels Bohr Institute, University of Copenhagen, Blegdamsvej 17, 2100 Copenhagen, Denmark}
}

\begin{document}
\begin{frontmatter}

\newpage

\maketitle


\vspace{-0.5cm}
\begin{abstract}
Gravitational waves from coalescing neutron stars encode information about nuclear matter at extreme densities, inaccessible by laboratory experiments. The late inspiral is influenced by the presence of tides, which depend on the neutron star equation of state.
Neutron star mergers are expected to often produce rapidly-rotating remnant neutron stars that emit gravitational waves.  These will provide clues to the extremely hot post-merger environment.
This signature of nuclear matter in gravitational waves contains most information in the $\unit[2-4]{kHz}$ frequency band, which is outside of the most sensitive band of current detectors.
We present the design concept and science case for a neutron star extreme matter observatory (NEMO): a gravitational-wave interferometer optimized to study nuclear physics with merging neutron stars.
The concept uses high circulating laser power, quantum squeezing and a detector topology specifically designed to achieve the high-frequency sensitivity necessary to probe nuclear matter using gravitational waves.  
Above one kHz, the proposed strain sensitivity is comparable to full third-generation detectors at a fraction of the cost.
Such sensitivity changes expected event rates for detection of post-merger remnants from approximately one per few decades with two A+ detectors to a few per year, and potentially allows for the first gravitational-wave observations of supernovae, isolated neutron stars, and other exotica.
\end{abstract}

\end{frontmatter}

\section{Introduction}
Gravitational-wave astronomy is reshaping our understanding of the Universe.
Recent breakthroughs include the detection of many gravitational-wave signals from binary black hole collisions~\citep{abbott18_O2catalog} leading to an enhanced understanding of their population properties~\citep{abbott18_O2population}, measurement of the Hubble parameter~\citep{abbott17_gw170817_Hubble,hotokezaka19}, unprecedented tests of Einstein's theory of General Relativity, including constraints on the speed of gravity~\citep{abbott17_gw170817_gwgrb} and hence the mass of the graviton~\citep{abbott17_gw170104,abbott19_gw170817_grtest}, to name a few.
Plans for building the next generation of observatories are afoot.  The United States National Science Foundation, Australian Research Council, and British government have financed an upgrade to Advanced LIGO (aLIGO) known as A+, which will increase the sensitivity of the current detectors by a factor of 2-3 dependent on the specific frequency of interest~\citep{Aplus}.
Research and development is ongoing for third-generation observatories, the Einstein Telescope~\citep{punturo2010einstein} and Cosmic Explorer~\citep{CosmicExplorer}: broadband instruments with capabilities of hearing black hole mergers out to the dawn of the Universe.

Third-generation observatories require substantial, global financial investments and significant technological development over many years.
To bridge the gap between A+ and full-scale, third-generation instruments, it is necessary to explore smaller-scale facilities that will not only produce significant astrophysical and fundamental physics outcomes, but will simultaneously drive technology development.
In this spirit, we introduce a Neutron star Extreme Matter Observatory (\NEMO): a dedicated high-frequency gravitational-wave interferometer designed to measure the fundamental properties of nuclear matter at extreme densities with gravitational waves.
We envision \NEMO{} as a specialized,  detector with optimum sensitivity in the kHz band operating as part of a heterogeneous network with two or more A+ sensitivity observatories.
The A+ observatories provide source localization while a \NEMO{} measures the imprint of extreme matter in gravitational-wave signals from binary neutron star mergers.
To maximise scientific impact, a \NEMO{} must exist simultaneously with 2.5-generation observatories, but before full-scale third-generation instruments are realised.

\begin{figure}[htb]
\centering
\includegraphics[width=1.\columnwidth]{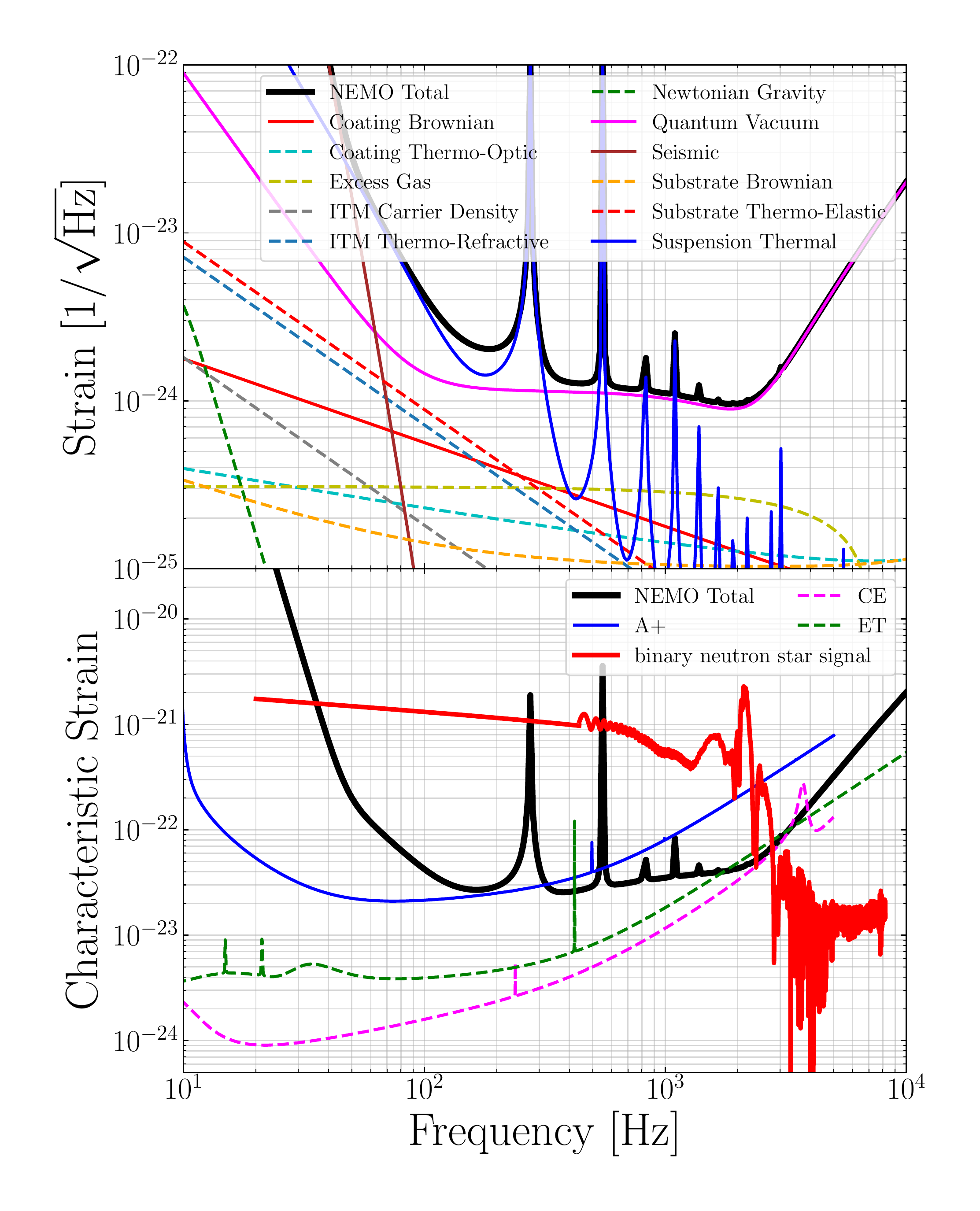}
\caption{Noise budget and indicative gravitational-wave signal from a binary neutron star collision.  Top panel: we show the amplitude spectral density of the various noise components that make up the total noise budget shown as the black curve.  Bottom panel: The black curve is the same total noise budget as the top panel, now shown as the noise amplitude $h_n=\sqrt{fS_n(f)}$, where $S_n(f)$ is the power-spectral density. This curve is shown in comparison to design sensitivity of A+ (blue), the Einstein Telescope (ET; green), and Cosmic Explorer (CE; pink).  Also shown in red is the predicted characteristic gravitational-wave strain $h_c$ for a typical binary neutron star inspiral, merger, and post-merger at 40 Mpc, where the latter are derived from numerical-relativity simulations.}
\label{fig:noisecurve}
\end{figure}

Neutron stars are an end state of stellar evolution.
They consist of the densest observable matter in the Universe, and are believed to consist of a superfluid, superconducting core of matter at supranuclear densities.
Such conditions are impossible to produce in the laboratory, and theoretical modelling of the matter requires extrapolation by many orders of magnitude beyond the point where nuclear physics is well understood.
As two neutron stars coalesce, their composition leaves an imprint on the gravitational waveform, which becomes increasingly important at higher frequencies $\sim\unit[0.5-4]{kHz}$.

Mergers produce remnants, some of which collapse to black holes, and some of which survive as long-lived, massive neutron stars.
Up to $\approx79\%$ of all binary neutron star mergers may produce massive neutron star remnants that emit strong gravitational-wave signatures~\citep{margalit19}.
The precise nature of the remnant is strongly dependent on the details of nuclear physics, which is encoded in the neutron star equation of state~\cite[e.g., see][and references therein]{bernuzzi19}.  
Measuring gravitational waves at these high frequencies therefore offers a window into the composition of neutron stars, not accessible with other astronomical observations or terrestrial experiments.
We show that detection rates of gravitational waves from post-merger remnants with a network of only two A+ observatories is between one per decade and one per century, while adding a \NEMO{} to the network increases this to more than one per year.

The technologies that will enable a \NEMO{} are key components for third-generation observatories. 
In order to reduce quantum shot noise, future detectors aim to employ aggressive squeezing (e.g., up to $\sim\unit[10]{dB}$).  
To enable increased circulating power, reduce scattering losses and thermal noise, future detectors may include cryogenic silicon test masses with high power $\unit[2]{\mu m}$ lasers as proposed in the Voyager design \citep{Adhikari_2020}.
A \NEMO{} observatory also provides technological development for Cosmic Explorer-like detectors while producing impactful science results on a shorter timescale.

Figure~\ref{fig:noisecurve} highlights both the key science case and the design principles for a \NEMO{} that are elucidated throughout the paper.  The top panel shows the strain sensitivity (amplitude spectral density $\sqrt{S_n(f)}$) and all underlying noise sources of the proposed detector.  This noise budget and the basic design principles of a \NEMO, including a detector schematic, are laid out in Sec.~\ref{sec:design}.  The bottom panel shows again the \NEMO{} noise budget in black, this time in terms of the noise amplitude $h_n=\sqrt{fS_n(f)}$, and a comparison with the design sensitivity curves of A+ (blue), Cosmic Explorer (pink), and the Einstein Telescope (green).  The sensitivity of a \NEMO{} is comparable to those third-generation instruments in the kHz regime.  Also shown in the bottom panel is an example signal one might expect from a binary neutron star merger at 40~Mpc, the same distance as the first binary neutron star merger detection GW170817.  Tidal effects during the inspiral become prominent around 500~Hz and above, while the post-merger signal is above 1~kHz.  We detail these key science deliverables and more in Sec.~\ref{sec:science}.   In Sec.~\ref{sec:conclusion}, we provide concluding remarks and sketch a path forward to a high-frequency detector within the international gravitational wave network. Finding \NEMO: a potential location for a \NEMO{} includes Australia, where a design concept called OzHF is eventually extended to a full-scale broadband Cosmic Explorer South.  For details see~\citet{bailes19}.

\section{Building \NEMO}\label{sec:design}

Simultaneously achieving high sensitivity at low ($\lesssim50$~Hz) and high ($\gtrsim1$~kHz) frequencies in a single detector is extremely challenging. There are two main reasons for this. First, the optical bandwidth of high-sensitivity kilometer-scale detectors is limited. Thus to achieve sensitivity peaked at $\approx2$~kHz requires a loss of optical sensitivity below $\approx500$~Hz. Second, the high circulating power required to improve high-frequency sensitivity introduces opto-mechanical instabilities whose control strategies can easily increase the noise in the low-frequency band. Detectors like the Einstein Telescope~\citep{ET} plan to limit low- and mid-band frequency noise sources such as thermal noise by operating at 20~K, which is not compatible with high circulating power. Broadband operation will then be achieved by building multiple detectors in a common subterranean vacuum envelope. In \NEMO, we only concentrate on the frequency regime above $\approx1$~kHz, sacrificing low-frequency sensitivity and thereby decreasing engineering challenges and cost. The low-frequency sensitivity required for sky localization will be achieved by the other detectors in the network.

\citet{martynov19} have shown that the optimal length of a detector with optimum sensitivity at 2kHz is 16~km. 
At this time, it is unlikely that the funds needed to build a dedicated high-frequency detector of this scale could be obtained, hence we have compromised to an arm length of 4~km which is also compatible with existing facilities. This arm length is sufficient to prevent displacement noise sources causing concern without being prohibitively expensive to build~\citep{miao18}. This reduction in arm length reduces the maximum sensitivity that can be obtained by about a factor of 2, which may in principal be recovered in a future upgrade using a folded interferometer as outlined in \citet{ballmer2013new}. 

\begin{figure}[htb]
\centering
\includegraphics[width=0.48\textwidth]{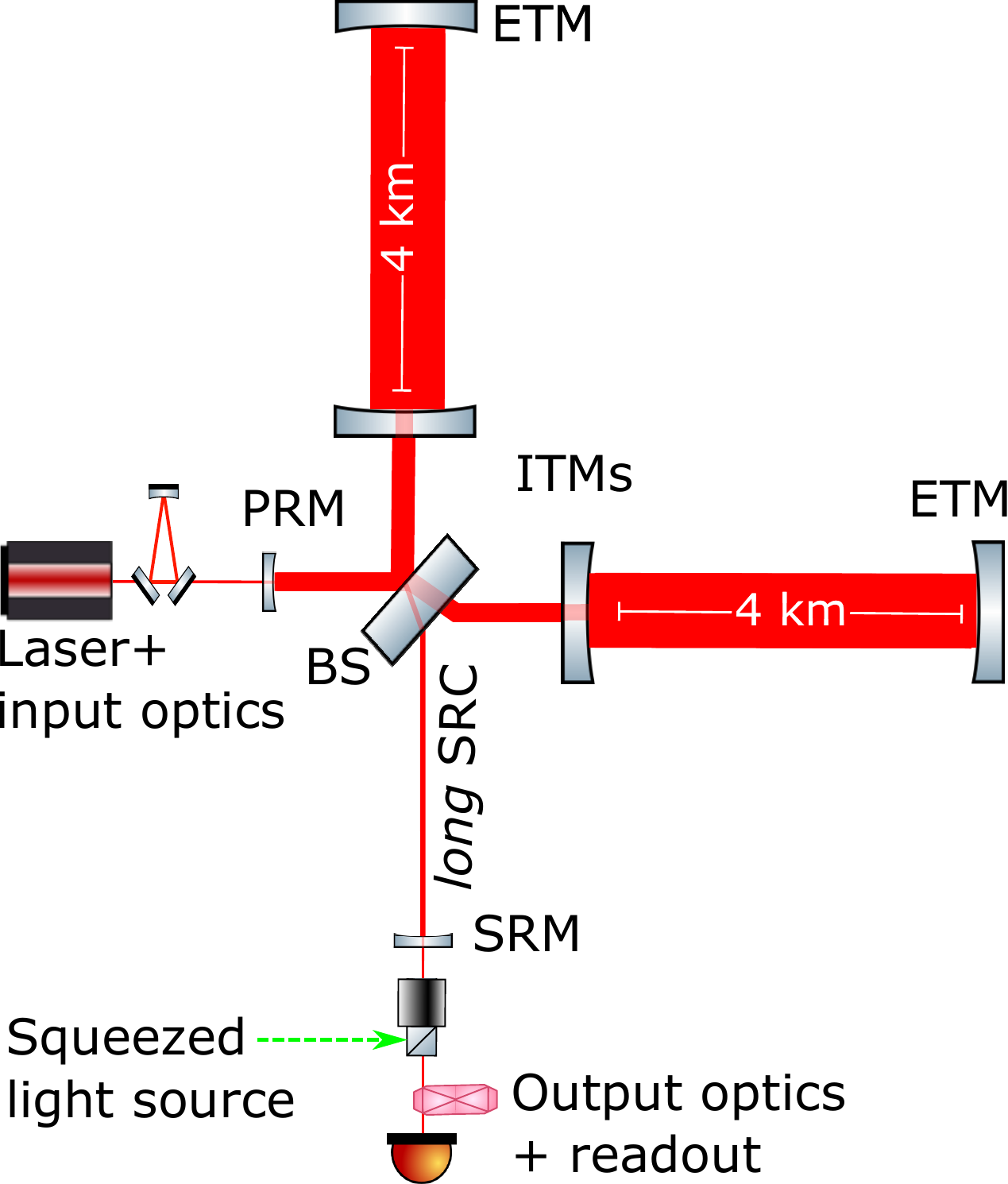}
\caption{Simplified optical topology of \NEMO. Folding of the recycling cavities, input and output optics, e.g. various mode cleaners, not shown for clarity. A summary of the design parameters is included in Tab.~\ref{tab:design}.  Acronyms in the figure are power recycling mirror (PRM), beam splitter (BS), input and end test mass (ITM and ETM, respectively) and Signal Recycling Cavity and Mirror (SRC and SRM, respectively).  
}
\label{detector_design}
\end{figure}

Our approach for achieving kilohertz sensitivity with a \NEMO{} that is comparable to third-generation gravitational wave observatories is outlined below. A simplified schematic of the inteferometer is illustrated in Fig. \ref{detector_design} and the design parameters are included in Table ~\ref{tab:design}.

The high-frequency sensitivity of interferometric gravitational-wave detectors is predominantly  limited by quantum phase noise, which is due to the quantum nature of light, and not displacement noise sources such as seismic and thermal. Increasing the circulating power within the detector reduces the impact of this quantum phase noise proportional to the inverse of the square root of the power~\citep{martynov19}.  Therefore, to maximize sensitivity, the circulating power in the arms must be as large as possible. This quantum phase noise source can also be reduced by injecting squeezed vacuum into the dark port~\citep{aasi2013enhanced}. As a baseline design, we choose 4.5~MW circulating power in the arms and inject 10~dB of squeezing which results in a 7~dB reduction in quantum noise. The circulating power and squeezing levels are chosen due to their feasibility given current technology constraints. If technology improves, both will be increased to further enhance the performance of the \NEMO{} detector.   

The quantum phase noise limited nature of high-frequency interferometers means that there are unlikely to be significant advantages in using exotic interferometer types such as speed meters~\citep{chen2003sagnac} or other Sagnac style interferometers~\citep{mizuno1997frequency}. For this reason, we choose a dual-recycled Michelson interferometer design with Fabry-Perot arm cavities, similar to current interferometric gravitational-wave detectors~\citep{aasi2015advanced,acernese2014advanced,aso2013interferometer}. However, there are some key differences targeted to maximise the sensitivity in the 1~kHz to 4~kHz signal band of interest.

The signal-recycling cavity and the arm cavity of the interferometer form a coupled cavity system which determines the overall bandwidth of the interferometer.  In order to maximise the sensitivity of a \NEMO{} in the 1~kHz to 4~kHz frequency band of interest, the length of the signal-recycling cavity and the transmission of the ITMs was optimised using numerical simulation tools PyKat and Finesse~\citep{Finesse,finesse1,brown2020pykat}. This resulted in the transmission of the input test mass being set to 1.4\% which resulted in 4.5~MW in the arm cavity and a signal-recycling cavity length of 354~m long. This `long' signal-recycling cavity displays the characteristic splitting of a coupled cavity system~\citep{DEM95,martynov19} around the interferometer carrier frequency where the gravitational-wave signal sidebands are resonantly enhanced. This splitting frequency is given by,
	\begin{equation}
	\mathrm{f_{sp}} = \frac{c\sqrt{\mathrm{{T_{ITM}}}}}{4 \pi \sqrt{\mathrm{L_{arm}}\mathrm{L_{src}}}},
	\label{eqn1}
	\end{equation} 
where $\mathrm{T_{ITM}}$ is the power transmissivity of the input test mass, $c$ is the speed of light, and $\mathrm{L_{arm}}$ and $\mathrm{L_{src}}$ are the lengths of the arm cavities and the signal-recycling cavity, respectively. The bandwidth $\gamma$ of this coupled cavity system depends on the transmission of the signal recycling mirror as well as the length of the signal-recycling cavity.
 
The same effect of enhanced sensitivity at certain frequencies can be obtained by detuning the signal-recycling cavity~\citep{SRCdetuned}. However, this configuration comes with technical challenges pertaining to the control of the interferometer~\citep{Ward_PhDThesis,Cahillane17}.

Cryogenically-cooled silicon test masses will be used to maximise the potential circulating-arm powers prior to adverse thermal distortions while also providing reduced coating thermal noise~\citep{Adhikari_2020} . At cryogenic temperatures, silicon exhibits high thermal conductivity and low thermal expansion~\citep{kim2018}. These test masses necessitate a departure from the $\unit[1.06]{\mu m}$ lasers used in current generation gravitational-wave detectors. 
A \NEMO{} operating with a 500~W, $\unit[2]{\mu m}$ single frequency diffraction limited laser is specified based on the silicon transmission window, the potential for reduced test-mass coating absorption~\citep{steinlechner2018}, and the relative technological maturity of prospective sources.
Thulium-doped fiber lasers are selected due to their intrinsic beam quality, demonstrated narrow linewidths at high power~\citep{goodno2009}, and potential for robust all-fibre architecture. While fiber-laser technology has been demonstrated at 200~W at $\unit[1]{\mu m}$ with the required intensity and frequency noise~\citep{buikema2019,wellmann2019}, this has not yet been demonstrated at $\unit[2]{\mu m}$. Recently, encouraging frequency and intensity noise levels have been demonstrated at low power using external cavity diode lasers~\citep{kapasi2020_2um-ecdl}, which constitutes an important stepping stone. Fortunately, the thresholds of non-linear processes that limit the maximum power level of single frequency fiber lasers increase faster than the wavelength squared~\citep{dawson2008}, implying it is likely that the 500~W of power required for a \NEMO{} will be demonstrated in the near future.

In order to obtain the sensitivity target of a \NEMO{} without further increasing the arm cavity power, introducing squeezed light is essential. For this purpose, we will inject 10~dB of frequency-independent squeezing into the interferometer, which will result in a quantum noise suppression of 7~dB. In the frequency range of 1~kHz to 5~kHz, only quantum phase noise suppression is required, so systems to rotate the suppression quadrature to quantum radiation pressure noise reduction such as filter cavities are not required. The 7~dB reduction in quantum noise from injected squeezing assumed here is realistic as a 6~dB reduction in quantum shot noise in a kilometer-scale detector~\citep{squeezeRecord_GEO2018} has already been demonstrated. Squeezing at $2~\mu$m has already been demonstrated~\citep{mansell2018observation,yap2019squeezed} and almost 12 dB of squeezing has been demonstrated at $1.06~\mu$m in the frequency band of interest here~\citep{stefszky2012balanced}. Custom photodetectors at $1.06~\mu$m 
have a quantum efficiency of 99.5\%~\citep{Vahlbruch15dB, Barsotti_2018} while at $2~\mu$m the best known efficiency for extended InGaAs to date is 74\%~\citep{mansell2018observation}. This has the implication that for 10 dB injected squeezing, the detected squeezing level stands lower than 4.5 dB~\citep{Barsotti_2018}. In order to improve the detected reduction in quantum noise, this quantum efficiency needs to be improved and is currently work in progress. There is however no fundamental reason as to why >90\% quantum efficiency photodiodes cannot be manufactured. 

\begin{table}[t]
    \centering
    \begin{tabular}{c c}
    
         Parameter & Value \\
         \hline
        \hline
         Laser Wavelength & $ 2~\rm{\mu m}$ \\
         Laser Power & 500 W \\
         Arm Length & 4 km \\
         Signal Recycling Cavity Length & 354~m \\
         Power Recycling Mirror Transmission & 3\% \\
         Input Test Mass (ITM) Transmission & 1.4\% \\
         End Test Mass (ETM) Transmission & 5~ppm \\
         Signal Recycling Mirror Transmission & 4.8\% \\
         SRC loss & 1400~ppm\\
         Power on Beamsplitter & 31.2~kW \\
         Arm Circulating Power & 4.5~MW \\
         Readout losses & 3000~ppm\\
         Reduction in quantum noise & 7~dB \\
         Test Mass Material & Silicon \\
         ITM Temperature & 150 K \\
         ETM Temperature & 123 K \\
         Test Mass Coating & AlGaAs/GaAs \\
         Test Mass Diameter & 45~cm \\
         Test Mass Thickness & 20~cm \\
         Test Mass Weight & 74.1~kg \\
         ITM Radius of Curvature & 1800~m \\
         ETM Radius of Curvature & 2500~m \\
         Beam Radius on ITM & 57.9~mm\\
         Beam Radius on ETM & 83.9~mm\\
         Suspension Fiber Length &	0.55~m \\
         Suspension Fiber Material & Steel \\
         Suspension Fibers per Test Mass &	4 \\
         Test Mass Cooling Method & Radiative \\
         Interferometer Configuration & Dual Recycled with  \\
          & Fabry Perot Arms \\
         \hline
    \end{tabular}
    \caption{Neutron Star Extreme Matter Observatory design parameters.  All recycling cavities are stable cavities. 
    }
    \label{tab:design}
\end{table}

The in-band noise performance can likely be achieved using a triple-stage suspension system similar to that of the aLIGO beam splitters and other auxiliary optics~\citep{robertson2010_triple-suspensions}. Additional seismic isolation requirements in the control band($\lesssim\unit[50]{Hz}$) should not be onerous and could be met by a simple active stage because of the increased actuation allowed on the test masses. The focus on the kilohertz band means that it is possible to use a steel suspension wires that are a reliable and proven suspension technology. Details of this are contained in a companion paper~\citep{Eichholz2020}. The \NEMO{} concept assumes only the last suspension stage to be cryogenic, both upper stages remain at room temperature. Further, to reduce the peak velocity of the optics in the pre-locked state, all the main optics are suspended by multi-stage suspension and isolation systems. 

The performance of second-generation detectors will be limited by coating thermal noise in the mid-band around 100 Hz once design sensitivity is reached. The order of magnitude improvement in quantum noise promised by high-frequency detectors promotes the coating thermal noise levels seen in current detectors to become a limitation at kHz frequencies, where it used to be of little concern.
The constraints on coatings are further increased due to the requirement that coating absorption is very low.

Several potential coating choices are being actively researched for use in cryogenic third-generation detectors, such as ion-beam sputtered amorphous oxides, non-oxides, and crystalline thin films of III-V semiconductor materials. Two promising crystalline coating options are epitaxially grown multilayers of the AlGaAs/GaAs system~\citep{cole2008algaas,cole2013algaas-nature} or the AlGaP/GaP system~\citep{lin2013algap,cumming2015algap,murray2017algap}. AlGaAs/GaAs coatings exhibit exceptionally low mechanical loss, optical absorption, and scatter. Of the possible amorphous coating materials, amorphous silicon ($\alpha$-Si) is particularly attractive due to its low mechanical loss and high refractive index. The pairing of $\alpha$-Si with silicon dioxide (SiO$_2$) as a low index material produces very thin coatings~\citep{steinlechner2016aSi,birney2018aSi}, which further benefits coating thermal noise due to its scaling with the square root of the coating thickness.

Currently, the choice of coatings is not clear cut as $\alpha$-Si/SiO$_2$ coatings have a thermal noise advantage but unacceptably high projected absorption losses of 20\,ppm~\citep{steinlechner2018}. On the other hand, AlGaAs/GaAs coatings suffer from elevated levels of thermooptic noise principally due to high thermo-refractive and thermal expansion coefficients. With careful coating design, the partially coherent thermo-refractive and thermo-elastic noise terms can cancel each other for an overall reduction in thermooptic noise~\citep{evans08_thermooptic,chalermsongsak16_photothermal-cancellation}. Another drawback of AlGaAs/GaAs coatings is that due to their lattice mismatch with silicon, they have to be grown on separate GaAs wafers and transferred onto the test masses. Steady progress is being made to upscale this technology with encouraging results~\citep{penn2019algaas}.

Both issues require time to be addressed, but the high absorption of $\alpha$-Si presents a more fundamental issue for a NEMO detector. While $\alpha$-Si is a broadly studied material for its use in photovoltaics, the cryogenic material properties of AlGaAs/GaAs are better documented in the literature, which makes it easier to reliably predict thermal noise levels. The temperature dependence of thermal noise in AlGaAs/GaAs coatings is more fully explored in~\citet{Eichholz2020}.  For the noise budget in Fig.~\ref{fig:noisecurve}, simple quarter-wave multi-layer AlGaAs/GaAs coatings that accomplish the ITM and ETM transmissions listed in Table~\ref{tab:design} were assumed.

AlGaAs/GaAs coatings are a new coating technology and their optimization and limitations have not been fully explored. It is worth pointing out that the titania-doped tantala/silica (TiO$_2$:Ta$_2$O$_5$/SiO$_2$) coatings of aLIGO and Advanced Virgo (AdVirgo) may be a suitable lower risk alternative coating for the NEMO detector despite their increase in mechanical loss towards cryogenic temperatures. Compared to AlGaAs/GaAs coatings, Brownian noise rises by roughly a factor of 3.8, but at the same time thermooptic noise is reduced by 35\,\% in the case of conservative quarter wave coatings. However, using aLIGO coatings would only result in a 15\,\% overall increase in detector noise of a NEMO because the thermal noise of AlGaAs/GaAs coatings is significantly below the quantum noise in the kHz band~\citep{Eichholz2020}. 

We choose to operate the input test masses of the interferometer at 150~K rather than the 123~K specified for other third-generation silicon designs. This will allow the high power absorbed in the test masses to be radiatively dissipated to the 77 K cooled shields that will surround the test masses.  We therefore do not require conductive cooling~\citep{Eichholz2020}, which can compromise the suspension thermal noise of the detector. The details of this design are outlined in a companion paper~\citep{Eichholz2020} and are summarized below.

Silicon at a temperature in the range of 120 to 150~K has a low thermal expansion coefficient and a very high thermal conductivity, resulting in low thermal distortion of mirror surfaces.
The thermooptic coefficient of silicon is higher than that of room-temperature silica that is used in the current detectors. However, the dramatically increased thermal conductivity of silicon means that the thermal lensing in the substrates will be reduced despite the greater absorption in silicon substrates compared with a similar room-temperature silica detector. The 10 ppm/cm assumed for the absorption of silicon substrates does not represent a fundamental limit on silicon absorption and we expect this to improve with time. If this does not improve then the thermal compensation system will need to work very well.  Calculations have suggested that two orders of magnitude of suppression of substrate thermal lensing caused by uniform substrate and coating absorption is possible~\citep{lawrence2003active} which should be sufficient to prevent thermal lensing from limiting the sensitivity of \NEMO{}. It should also be noted that we calculate our choice of a $2 \mu m$ cryogenic silicon based detector reduces this effect by a factor 6 c.f a room temperature $1 \mu m$ silica based detector.

Point absorbers on the high reflectivity surfaces of the test masses have caused significant local distortions of these surfaces in the aLIGO detector~\citep{LIGO_O3_DET}. The impact of point absorbers will be reduced by a factor of over 300 for a silicon interferometer operating at 150~K compared with a silica interferometer operating at room temperature~\citep{Eichholz2020}.

To maximise the surface area for radiative heat transfer, we assume a mirror diameter of 45~cm, which is projected to become available in the form of single crystal cylindrical silicon boules grown by a magnetically-assisted Czochralski method (m-Cz) for semiconductor applications~\citep{lin2008_silicon}. Absorption levels of 10~ppm/cm have been demonstrated in m-Cz silicon~\citep{Adhikari_2020}. A thickness of 20~cm results in a total mass of 74.1~kg. A black body of equivalent dimensions, held at 123~K, thermally radiates a total power of 7.8~W into its environment. At 150~K, this increases to 17.2~W. We assume that a cooling rate of about 70\% of these values can be achieved, as shown by detailed finite element simulations for Voyager~\citep{Adhikari_2020}, resulting in 5.5~W and 12.1~W, respectively.

A heat load of 4.5~W on the test masses is expected from a residual 1~ppm absorption of the high-reflectively coatings. Heating due to the transmitted light in the end test masses is negligible, however, with about 31~kW incident on the beam splitter, and each beam performing a double-pass through its respective input test mass, the bulk heating power becomes 0.31~W/cm, for a total of 6.2~W. We therefore select an elevated temperature of 150~K for the ITMs, while the ETMs remain at 123~K. At these temperatures, the radiative cooling rate provides a margin of more than 1~W to the budgeted beam heating. For more details on this elevated temperature operation, see~\citet{Eichholz2020}. Summarizing, we can state that radiatively cooling the input test mass, considering the heat load by absorption, needs an elevated temperature to increase the thermal gradient between test mass and cold shield. We model a trade-off between this and increased thermal lensing and low-frequency thermal noise to give an optimum temperature of 150 K. 

The beam splitter material choice is still an open question. The 31~kW incident on the beam splitter will result in significant astigmatic thermal lensing, even when the absorption in the substrate is low. In this situation, as with the GEO600 detector~\citep{Wittel18_BSComp} the beam splitter will need to be compensated. Different schemes to provide this compensation are actively being investigated.  

Experience with current gravitational-wave detectors has shown that opto-mechanical instabilities arise when operating with high circulating powers, such as parametric instabilities~\citep{evans2015observation}, and angular misalignment~\citep{sidles2006optical,hirose2010angular}. The high circulating power inside the arm cavities could make a \NEMO{} quite sensitive to opto-mechanical instabilities. However, this is where the dedicated high-frequency nature of the detector really comes into its own. In the case of angular instabilities, the bandwidth of the angular control loops can be significantly increased beyond what can be used for broadband detectors.  Modeling has shown even at 5~MW of circulating power the coupled opto-mechanical tilt modes of the \NEMO{} arm cavities will not exceed 15~Hz. 
We estimate that these tilt modes can be controlled with angular control bandwidth of $\sim$3 times the modified mode frequency, with sufficient noise suppression ($>$60~dB) at $\sim$10 times the modified mode frequency~\citep{Barsotti2010CQG, Adhikari_2020}. Hence, the noise injected by the required angular control loops should be insignificant for frequencies above $\approx$1 kHz. 

Parametric instability was first observed at aLIGO with about 40~kW~\citep{Evans2015_PI} circulating power in the arms, while AdVirgo~\citep{acernese2014advanced} did not observe parametric instabilities with circulating power of around 100~kW in 2019. This is indicative of the sensitivity of parametric instability to cavity and test-mass parameters, detailed models are required for an accurate prediction of parametric instability.  This detailed analysis is currently being performed~\citep{Juntao2020}.  However, rough estimation can be performed by assuming a scaling of aLIGO parameters and related scaling of the severity of parametric instabilities described in~\citet{Braginsky01}.

Parametric-instability severity scales proportional to circulating power and optical quality factors, and inversely proportional to mirror mass and mechanical eigen frequencies.  The power will be 112.5 times higher than where aLIGO first observed parametric instability, the optical quality factors will be 1.35 times higher, the mirror mass will be 1.8 times heavier, and the lowest mechanical frequency is 1.08 times higher. If other parameters are considered unchanged, parametric instability is expected to be 98 times worse than aLIGO with 40~kW circulating power from this scaling argument.  However, thermal tuning~\citep{Zhao_2006_Therm,Hardwick2020_DTC} allowed optical power to be increased by a factor of 4.3 at aLIGO.  Resonant mass dampers attached to the test masses have been demonstrated to reduce mechanical mode quality factors by 10 to 100~\citep{Biscans2019_AMD}, introducing negligible thermal noise and electrostatic actuation has been demonstrated to reduce parametric gain by a factor of 13 and is inferred to be strong enough to reduce mode quality factors by 1000s~\citep{Blair_2017_ESDPI}.  This leads us to believe that with proper consideration of parametric instability and its mitigation it may be controlled in a \NEMO.
 
The noise budget for NEMO is illustrated in Fig. \ref{fig:noisecurve}. Assuming that the vacuum envelope will have similar or better performance than current detectors, the sensitivity of \NEMO{} will be limited by quantum noise above 500~Hz. All other noise sources are a factor of 5 below this. The performance of a \NEMO~is compared to A+, Cosmic Explorer and Einstein Telescope in the bottom panel of Fig.~\ref{fig:noisecurve} (bottom) which clearly illustrates that \NEMO~has comparable performance to the third generation detectors around 2~kHz.

\section{Scientific deliverables}\label{sec:science}

To motivate the science case for a \NEMO, we discuss physics encoded in the kilohertz gravitational-wave emission during both the inspiral and post-merger phases of a binary neutron star merger.
These two phases probe different temperature regimes of the neutron star equation of state.
During inspiral, neutron stars are relatively cold, with temperature $T\ll\unit[10^{9}]{K}$, having had sufficient time to cool since birth.  
Under such conditions, the temperature does not significantly affect internal physical structure that determines bulk stellar quantities such as the stellar radius.  
Temperatures during merger can reach as high as $T\sim\unit[10^{11}]{K}$~\citep[e.g.,][]{Baiotti2008,foucart16}, and can therefore affect the equation of state.

\subsection{The physics of cold neutron stars}

For cold neutron stars in the pre-merger phase, the tidal deformation of the individual components is imprinted in the gravitational-wave emission.  The tidal deformation is dependent on the equation of state, and is parameterized by the ``combined dimensionless tidal deformability'' $\tilde\Lambda$, given by:
\begin{equation}
    \tilde\Lambda \equiv \frac{16}{13}\frac{(m_1+12m_2)m_1^4\Lambda_1+(m_2+12m_1)m_2^4\Lambda_2}{(m_1+m_2)^5}.
\end{equation}
Here $m_1$ and $m_2$ are the masses of the component neutron stars, and $\Lambda_1$ and $\Lambda_2$ are the tidal deformabilities of each neutron star, defined as
\begin{equation}
    \Lambda_i\equiv\frac{2k_{2,i}}{3}\left(\frac{c^2R_i}{Gm_i}\right)^5,
\end{equation}
where $R$ is the radius, and $k_{2}$ is the second Love number, which measures the rigidity of the neutron star.
Gravitational-wave astronomers measure $\tilde\Lambda$ because it is the leading-order correction to gravitational waveforms due to tides.
For a fixed mass, both $R$ and $k_2$ are determined by the neutron star equation of state.
Small values of $\Lambda$ imply soft equations of state, corresponding to small, compact neutron stars.  Large values of $\Lambda$ imply stiff equations of state, where neutron stars are large and comparatively fluffy.
Black holes have $k_2=0$, implying the tidal deformability also vanishes.

A key goal in nuclear astrophysics is to measure the tidal deformability as a function of neutron star mass.  
These tidal effects become increasingly important when the two neutrons stars are close to one another, which occurs late in coalescence and therefore at kilohertz gravitational-wave frequencies.
In the bottom panel of Fig.~\ref{fig:noisecurve}, we plot the \NEMO{} (black) and A+ (blue) noise amplitude curves $h_n(f)=\sqrt{fS_n(f)}$, where $S_n(f)$ is the detector power spectral density, alongside the gravitational-wave characteristic strain $h_c=2f\tilde{h}(f)$ from the inspiral and postmerger phase of an equal-mass binary neutron star coalescence at \unit[40]{Mpc} (red).  With these quantities, the expected signal-to-noise ratio $\rho$ is simply~\citep[e.g.,][]{moore15}
\begin{align}
    \rho^2=\int_{-\infty}^\infty d\ln f\left(\frac{h_c(f)}{h_{n}(f)}\right)^2.
\end{align}
Tidal effects become important at frequencies~$\gtrsim\unit[400]{Hz}$~\citep{harry18}; at that point, the gravitational waveforms describing a binary black hole system and a binary neutron star system begin to dephase.  A \NEMO{} is designed to be sensitive to the physics of this late inspiral phase.

To study the sensitivity with which a \NEMO{} can measure tidal deformability, we perform a Monte Carlo study in which we inject binary neutron star inspiral signals into simulated noise from two different detector networks.
Network I consists of two A+ detectors located at Hanford and Livingston, and Network II is a three-detector network that adds a \NEMO{} observatory.  We locate the third detector in Gingin, near Perth in Australia.

We assume the population of mergers is distributed uniformly in co-moving volume, with a binary neutron star merger rate given by the mean merger rate inferred in~\citet{GW190425}. In just over six months of observation, this corresponds to 44 detected binary neutron star merger signals with matched-filter signal-to-noise ratio $\rho_{\rm mf}>20$ with Network I, and 61 such detected signals with Network II.  We choose $\rho_{\rm mf}>20$ as signals weaker than this do not contribute appreciably to the cumulative inference of the equation of state~\citep{hernandez19}.

For simplicity, we further assume the chirp mass of these systems is uniformly distributed between 1 and $\unit[1.74]{M_\odot}$, and that all systems are equal-mass, non-spinning binary mergers.  We do not expect these assumptions to change our inference of the equation of state.  Our injections are performed using an SLy equation of state~\citep{douchin01}, and we calculate the uncertainty on the masses, tidal deformability, time of coalescence, and phase using a Fisher matrix approximation~\citep{martynov19}; we ignore uncertainties on other parameters as they do not correlate with the equation of state~\citep[e.g., see][and references therein]{abbott17_gw170817_detection}.  We reconstruct the equation of state following the procedure outlined in~\citet{lackey15} and~\citet{hernandez19}.

\begin{figure}[htb]
\centering
\includegraphics[width=0.5\textwidth]{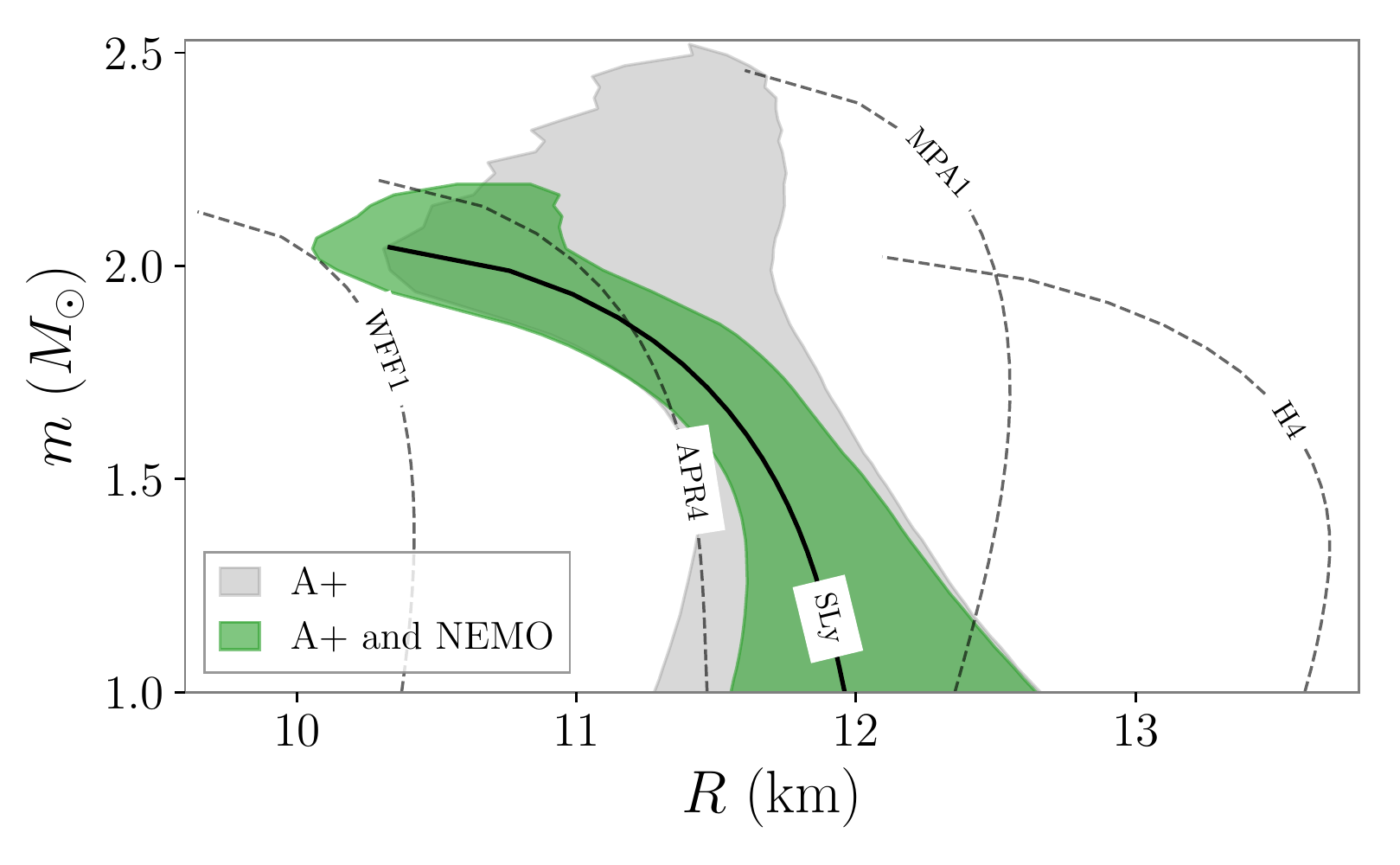}
\caption{
Reconstruction of the neutron-star mass radius relation using simulated data from $40$ binary neutron star mergers.
The true relation is shown with the solid black line, labelled SLy, while alternative equations of state are shown with dashed black curves.
The grey contour shows the 90\% credible interval obtained using a network of two A+ interferometers, while the green shows the same interval obtained when a \NEMO{} is added to the network, assuming approximately six months of operation.  In this example, from all of the equations of state shown, all but SLy equation are ruled out.
}
\label{fig:mass_radius}
\end{figure}

In Fig.~\ref{fig:mass_radius}, we show a reconstruction of the mass-radius relation using the above prescription.
The grey and green contours show respectively the 90\% credible interval obtained using Network I, and Network II (i.e., adding a \NEMO), and the SLy equation of state that we use for the injection is shown as the solid black curve.  We also show other a set of other indicative equations of states commonly used in gravitational-wave analyses~\cite[e.g.,][]{abbott17_gw170817_detection, GW190425}.   The inclusion of a \NEMO{} provides a significant improvement in the accuracy with which the equation of state can be measured, due to the improved high-frequency sensitivity of the network attributed to the addition of NEMO.  For example, we see that the constraints on the equation of state are significantly less than one kilometer at a fixed mass.  Figure~\ref{fig:mass_radius} also shows the posterior distributions are well-enough constrained to rule out a large number of equations of state.  In this figure, one would confidently rule out all equations of state, including APR4 with Network II, but not necessarily with Network I.  

Our equation of state posterior distributions also imply we can infer the maximum neutron star mass allowed by that equation of state.  In the example presented here, we are able to constrain the maximum mass with an accuracy of $\approx\unit[0.3]{M_\odot}$ at 90\% confidence with Network I, which improves to $\approx\unit[0.15]{M_\odot}$ with the inclusion of a NEMO.  This method for constraining the important maximum neutron star mass is complementary to the myriad of other direct and indirect methods in the literature~\cite[e.g.,][]{margalit17, ruiz18, alsing18,sarin20,chatziioannou20, landry20}

In general, equation of state constraints such as those presented in this section and highlighted with Fig.~\ref{fig:mass_radius}, are complementary to those using other methods such as x-ray and radio observations of isolated and accreting neutron stars~\cite[see][for a review]{lattimer07}, and observations of post-merger remnants (see below).  Each of these methods relies on different modelling assumptions and/or probes different regions of the equation-of-state parameter space.

\subsection{The physics of hot neutron stars}
Following the merger of two neutron stars, a new compact object is created.  Depending on the remnant mass, this compact object can be a black hole or a massive neutron star.  
In the former case, gravitational-wave emission is difficult to observe because of the relatively short damping time and high frequency $\gtrsim\unit[6]{kHz}$~\citep{echeverria89}. 
However, if a neutron star survives the merger, gravitational waves can be emitted at frequencies of $\sim\unit[1-4]{kHz}$ for up to hundreds of milliseconds~\citep{Baiotti2008,Shibata2006}.
The spectral content of the post-merger gravitational waves contains information about the neutron star equation of state~\citep{takami15}.
Following merger, the temperature becomes an important equation-of-state parameter.
For example, temperature-dependent phase transitions may occur in the core of post-merger neutron stars.  Measuring gravitational waves from the inspiral and post-merger phase could provide a unique opportunity to identify phase transitions from hadronic matter to deconfined quark matter~\citep{Bauswein2019}.
Furthermore, as the remnant is supported by differential rotation the resulting neutron star in the post-merger phase has a higher density than the component neutron stars from the pre-merger phase.  Thus gravitational-wave emission from the post-merger phase affords the opportunity to probe the equation of state in a different density regime.

The precise signal morphology of neutron star post-merger gravitational waves remains unknown. However, numerical simulations have shown that the spectra of the emission from the nascent neutron star contain a characteristic peak frequency, approximately related to the fundamental quadrupolar mode of that neutron star, and lower-frequency peaks~\cite[e.g.,][and references therein]{bauswein15}.

Using an algorithm that reconstructs gravitational-wave signals as a sum of sine-Gaussian wavelets called \textsc{BayesWave}~\citep{BayesWave}, the post-merger waveform can be reconstructed with minimal assumptions on the exact morphology of the signal.  From this reconstruction, it is possible to produce posterior distributions of the characteristic peak frequency.
For signals where the post-merger matched filter signal-to-noise ratio $\rho_\text{mf}\gtrsim5$, the peak frequency can be constrained to within tens of Hz~\citep{Chatziioannou,TorresRivas}.

\begin{figure}[htb]
\centering
\includegraphics[width=0.5\textwidth]{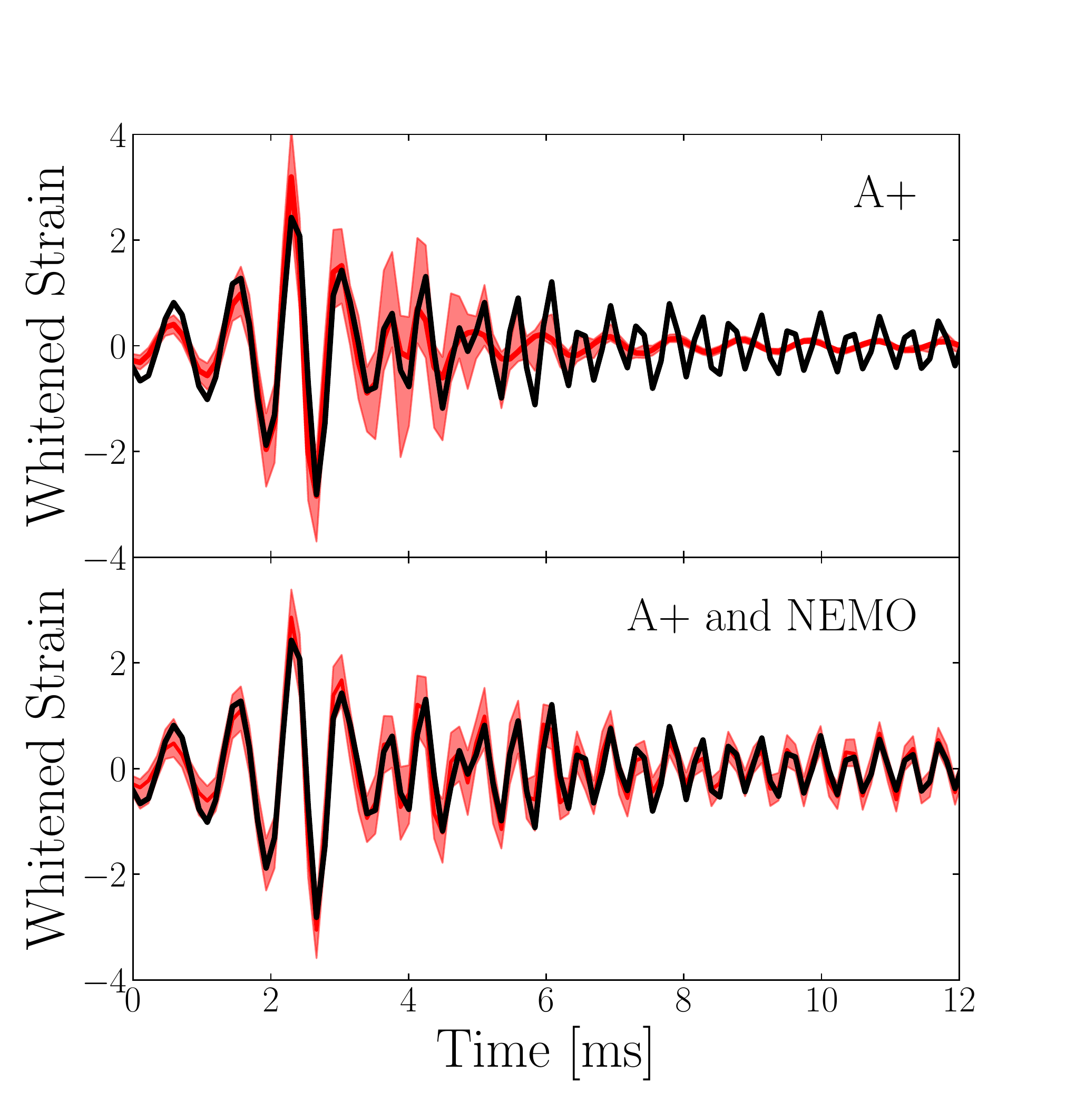}
\caption{Gravitational-wave reconstruction of a post-merger signal with and without a \NEMO.  We inject the same numerical-relativity post-merger waveform as Fig.~\ref{fig:noisecurve} into a detector network consisting of two A+ detectors (top panel), and two A+ detectors and a \NEMO{} detector (bottom panel).  The black curve shows the injected signal, while the shaded regions show the 90\% confidence interval reconstruction.  Without a \NEMO, the signal reconstruction fails to track the phase of the signal, whereas with a \NEMO{}, the gravitational-wave phase is correctly tracked throughout the signal duration.}
\label{fig:post-merger}
\end{figure}

In Fig.~\ref{fig:post-merger}, we show an example of a reconstructed post-merger signal for an event like GW170817 obtained using the \textsc{BayesWave} algorithm following~\citet{Chatziioannou}.
The top panel shows the 90\% credible interval obtained with Network I (two A+ observatories) and the bottom panel shows the 90\% credible interval obtained with Network II (adding a \NEMO).
Qualitatively, without a \NEMO{} only the first couple of milliseconds of the post-merger signal are reconstructed, while the reconstruction with a \NEMO{} correctly tracks the signal for $\gtrsim10$ ms. 
In the top panel of Fig.~\ref{fig:post-merger}, the reconstruction is consistent with zero signal---the immediate post-merger signal (from approximately $t=0$ to 4~ms) is reconstructed with short-lived wavelets, while the wavelets that fit the inspiral are relatively long-lived and well constrained.  The small extent of the 90\% credible interval after $\approx5$~ms is therefore an artefact of these basis functions, where tight constraint on the pre-merger signal leaks into post-merger region.  The estimation of the peak frequency only uses wavelets with central times after the merger, implying these small post-merger artefacts do not affect the peak-frequency posterior, nor the inferred physics of the post-merger remnant; for details, see~\citet{Chatziioannou}.
Using Network I, the characteristic peak frequency is mostly unconstrained with the 90\% credible intervals covering most of the prior range.  However when adding a \NEMO{} as in Network II, the characteristic peak frequency is constrained to a few tens of Hertz.  In other words, the addition of a \NEMO{} allows for a stringent measurement of the hot equation of state, whereas no information is gained about the post-merger remnant in the case of the two A+ detectors.

\begin{figure}[htb]
\centering
\includegraphics[width=0.5\textwidth]{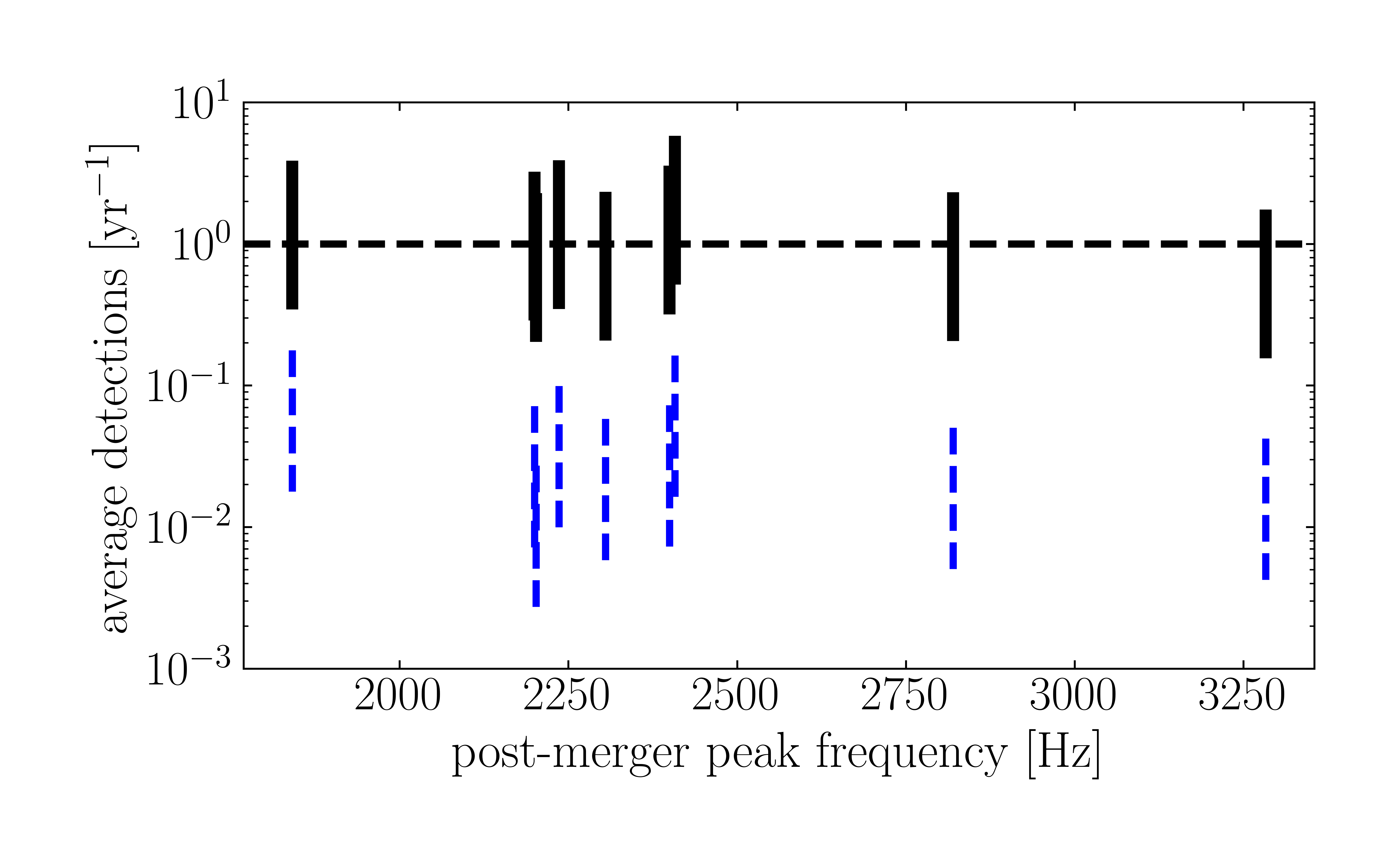}
\caption{The expected number of post-merger signals per year with matched-filter signal-to-noise ratio $\rho_\text{mf}>5$ as a function of peak gravitational-wave frequency.
In dashed blue, we show the number of detections for Network I with two A+ observatories while black indicates the number of detections for Network II when we add a \NEMO{} to make a three-detector network.
The length of the vertical lines shows the 90\% credible interval owing to uncertainty from the binary neutron star merger rate~\citep{GW190425}.}
\label{fig:ndet}
\end{figure}

In order to showcase the advantage gained by a \NEMO, we plot in Fig.~\ref{fig:ndet} the number of expected post-merger events per year for Network I with two A+ observatories (dashed blue) and Network II, which adds a \NEMO{} (solid black).
We inject post-merger gravitational waveforms from numerical-relativity simulations for a variety of equations of state.  
We calculate the matched-filter signal to noise for each signal injection for a realistic distribution of source distances, orientations, etc, and then calculate the average number of detections (defined as having $\rho_{\rm mf}>5$) per year in either network.
The results are plotted as a function of peak frequency, which depends on the equation of state.
The length of each line indicates the 90\% credible interval due to uncertainty in the binary neutron star merger rate~\citep{GW190425}.

Figure~\ref{fig:ndet} clearly shows that the average number of detections per year for a network of A+ interferometers \textit{not} including a \NEMO{} is significantly below one.  Put another way, one would have to wait potentially many tens of years for a first post-merger detection without a \NEMO.  That number increases to an average of about one detection per year with a \NEMO, ranging anywhere from one every few years to a few per year.  We emphasise that the uncertainty here encodes our uncertainty in the binary neutron star merger rate.

\subsection{Other Science}
Neutron-star science is the key science driver for a \NEMO.
Since the natural timescale for neutron-star physics is ${\cal O}(\unit[1]{ms})$, the frequency of gravitational waves from binary neutron star mergers is well-matched to a \NEMO.
Recent observations of binary neutron star mergers~\citep{abbott17_gw170817_detection, GW190425} make neutron-star science low-risk because there is no doubt that binary neutron stars merge frequently enough for \NEMO{} science.
The direct measurement of the effects of matter in binary neutron stars will facilitate additional science, for example, breaking degeneracies in measurements of the Hubble flow~\citep{messenger12, calderon20}, helping to distinguish between neutron stars and black holes~\cite[e.g.,][]{fasano20}, and any potential cosmological effects on the equation of state~\cite[e.g.,][]{haster20}.  

The operation of a \NEMO{} in a heterogeneous gravitational-wave network with two A+ interferometers will allow an unprecedented view into the hearts of short gamma-ray bursts.  Low-frequency A+ sensitivity will see negative-latency triggers for electromagnetic telescopes---that is, telescopes will receive alerts \textit{before} the two neutron stars merge---allowing early-time, multi-wavelength observations of the prompt emission, afterglow, and kilonovae~\cite[e.g.,][]{metzger18,james19}.  That information, together with gravitational-wave observations elucidating the nature of the remnant~\cite[e.g.,][]{shibata19} and time it takes for the remnant to collapse to a black hole~\citep{lucca19, easter20}, will be as important to our understanding of gamma-ray burst physics as the first multimessenger gravitational-wave observation GW170817/GRB170817A.  
The precise nature of the remnant of GW170817 is not known~\cite[e.g.,][]{ai20}, in large part due to the non-detection of post-merger gravitational waves~\citep{abbott_gw170817_postmerger1, abbott_gw170817_postmerger2}.
The joint electromagnetic and gravitational-wave detection of GW170817-like events with the addition of a \NEMO{} will enable significant further insight into gamma-ray burst physics.
For example, the delay time between the collapse and the prompt emission will drive studies into the jet-launching mechanism~\cite[e.g.,][]{zhang19,beniamini20} that is currently ill-understood~\citep{zhang18}, and the existence and lifetime of the remnant will reveal the impact of neutrino radiation on heavy-element formation through the rapid neutron-capture process in kilonovae~\citep{metzger14,martin15,fernandez19,kawaguchi20}.  

Neutron star-black hole mergers are also a target for a \NEMO, where the primary science case is again to measure the tidal effects of a neutron star through the inspiral and merger phase.
In general, neutron star-black hole binaries are a less sensitive probe of the equation of state than binary neutron star mergers (e.g., see~\citet{prayush17}, cf.~\citet{lackey15, hernandez19} for binary neutron stars).
The unknown rate estimates for neutron star black hole binaries imply it is difficult to estimate the frequency of detections, and therefore the potential science output. However, this situation may rapidly change with numerous neutron star-black hole candidates identified made during the third observing run of aLIGO and AdVirgo\footnote{\url{https://gracedb.ligo.org/superevents/public/O3/}}.

Additional sources {\em may} be within the reach of a \NEMO, for example, supernovae~\cite[e.g.,][]{powell19}, quasi-monochromatic signals from rotating neutron stars~\cite[e.g.,][]{lasky15,riles17}, or more speculatively, superradiance from axion clouds~\citep{yoshino14}.
However the detectability (and/or existence) of these sources is more speculative, and hence the great scientific impact of these targets must be tempered with theoretical uncertainty.
Other sources such as binary black holes and the stochastic background are more easily studied at lower frequencies; a \NEMO{} can detect them, but no better than broadband observatories such as A+.

By expanding the observing band of gravitational-wave networks, a \NEMO{} will explore a new region of parameter space.
History suggests that opening a new window on the Universe often yields unexpected discoveries; gamma-ray bursts are a great example.
While a \NEMO{} may detect something unexpected, we can be confident that it will measure the properties of matter effects in neutron stars.

\section{Conclusion}\label{sec:conclusion}
We present the technology requirements and key science drivers for an Extreme Matter Observatory: a kilohertz gravitational-wave observatory optimized to study nuclear physics with merging neutron stars.  A \NEMO{} utilises high-circulating laser power and quantum squeezing to achieve necessary high-frequency noise, while sacrificing difficult and costly low-frequency sensitivity.  Reaching a strain sensitivity of $\approx\unit[10^{-24}]{1/\sqrt{Hz}}$ in the $\sim$1--$\unit[3]{kHz}$ regime allows gravitational waves from the post-merger remnant of a binary neutron star collision to be detected with sufficient regularity.  Such a \NEMO{} should operate simultaneously with the A+ network which drives the sky localisation of sources and enables rapid electromagnetic identification of neutron star merger counterparts.  The combination of electromagnetic observations, such as those achieved for GW170817, together with precision gravitational-wave observations of the inspiral, merger, \textrm{and} post-merger remnant will provide unprecedented insight into both the \textrm{hot} and cold equations of state of nuclear matter at supranuclear densities.

The timescale for the development and construction of a \NEMO{} is driven on two fronts.  First, the science is maximised in a heterogeneous network of interferometers, such that the broadband A+ instruments realise the sky localisation of sources.  Second, that \NEMO{} is a key technology driver for full-scale third-generation instruments implies it must operate prior to The Cosmic Explorer and Einstein Telescope. Realistically, such a \NEMO{} could be operational in the late 2020s and early 2030s, giving sufficient time for co-operations with the A+ network while impacting technology development for third-generation detectors.  Such a proposal relies on engineering and detector design studies to be funded and implemented soon.  Preliminary investigations show that a \NEMO{} costs on the order of \$50 to \$100 M, a fraction of the $\sim$billion-dollar budget required for third-generation broadband instruments.

The location of a \NEMO{} is less critical than that of broadband detectors, where the network relies on long baselines to increase sky localization accuracy.  One suitable location includes Australia, where the OzHF concept~\citep{bailes19} sees the four-kilometer \NEMO{} eventually extended into a $10$s of km-scale, broadband Cosmic Explorer South; the need for which has been identified by the Gravitational Wave International Committee to, for example, enable precision localisation of all merging stellar-mass binary black holes throughout the Universe. 

\begin{acknowledgements}
We are grateful to Matt Evans and the anonymous referee for valuable comments on the manuscript.  This work was supported through Australian Research Council (ARC) Centre of Excellence CE170100004, ARC Future Fellowships FT150100281, FT160100112, and FT190100574, ARC Discovery Project DP180103155, and the Direct Grant, Project 4053406, from the Research Committee of the Chinese University of Hong Kong.
\end{acknowledgements}

\bibliographystyle{pasa-mnras}
\bibliography{bibfile}

\end{document}